\documentclass[aps,amsmath,amssymb,prd,preprintnumbers,showpacs]{revtex4}

\usepackage{amsmath}
\usepackage{bm}
\usepackage[dvips]{graphicx}
\usepackage{color}

\newcommand{\be}{\begin{equation}}

\newcommand{\ee}{\end{equation}}
\newcommand{\bea}{\begin{eqnarray}}
\newcommand{\eea}{\end{eqnarray}}

\begin{document}
\preprint{BARI-TH 564/07}

\author{R.~Gatto}
\affiliation{D\'epart. de Physique Th\'eorique, Universit\'e de
Gen\`eve, CH-1211 Gen\`eve 4, Suisse}

\author{M.~Ruggieri}
\affiliation{Dipartimento di Fisica, Universit\`a di Bari, I-70126
Bari, Italia}\affiliation{I.N.F.N., Sezione di Bari, I-70126 Bari,
Italia}

\date{\today}

\title{On the ground state of gapless
two flavor color superconductors}

\begin{abstract}
This paper is devoted to the study of some aspects of the
instability of two flavor color superconductive quark matter. We
find that, beside color condensates, the Goldstone boson related to
the breaking of $U(1)_A$ suffers of a velocity instability. We
relate this wrong sign problem, which implies the existence of a
Goldstone current in the ground state or of gluonic condensation, to
the negative squared Meissner mass of the $8^{th}$ gluon in the g2SC
phase. Moreover we investigate the Meissner masses of the gluons and
the squared velocity of the Goldstone in the multiple plane wave
LOFF states, arguing that in such phases both the chromo-magnetic
instability and the velocity instability are most probably removed.
We also do not expect Higgs instability in such multiple plane wave
LOFF, at least when one considers fluctuations with small momenta.
The true vacuum of gapless two flavor superconductors is thus
expected to be a multiple plane wave LOFF state.

\end{abstract}

\maketitle

\section{Introduction}

It is by now well accepted that quark matter at high densities and
low temperatures finds itself in a color-superconductive
state~\cite{Collins:1974ky}. The ground state at asymptotically
large quark densities is the CFL (color-flavor-locked)
phase~\cite{Alford:1998mk}. Reviews of high density QCD are given in
\cite{Rajagopal:2000wf,Alford:2001dt,Nardulli:2002ma,Reddy:2002ri,Schafer:2003vz,
Rischke:2003mt,Buballa:2003qv,Huang:2004ik,Shovkovy:2004me,Alford:2006fw}.
It is also admitted however that, if the cores of some compact stars
contain quark matter, such matter should be in a different
pre-asymptotic phase. The reason is that the expected values of the
chemical potential within such cores would be low enough to make non
negligible the strange quark mass, requiring a different type of
pairing than for CFL. The identification of such a phase, respecting
electric and color neutrality and weak equilibrium, is an active
subject of research at this time, and some solutions have been
proposed. Homogeneous and non homogenous fermion condensates have
both been examined, the latter case seeming at this time to be
better resistant to pathologies. The homogeneous choice leads to
g2SC in the case of two flavors \cite{Shovkovy:2003uu},
corresponding to analogous situations in condensed
matter~\cite{Gubankova:2003uj}. For three flavors it leads to gCFL
\cite{Alford:2003fq,Ruster:2004eg,Fukushima:2004zq,Abuki:2004zk,Ruster:2005jc,Blaschke:2005uj}.

Both g2SC and gCFL are gapless phases since the fermion dispersion
laws may have nodes for some value of the quasi-quark momenta (in
ordinary superconductors the fermion quasi-particles have node-less
dispersion laws). As a consequence there exist regions in the phase
space where fermion pairings become impossible. The pathology with
these phases is however a chromo-magnetic instability, both for two
flavors~\cite{Huang:2004bg}, and for three
flavors~\cite{Casalbuoni:2004tb}. Phases with inhomogeneous quark
condensates~\cite{Alford:2000ze,Casalbuoni:2001gt,Bowers:2001ip,
Leibovich:2001xr,Kundu:2001tt,Casalbuoni:2002pa,Bowers:2002xr,Casalbuoni:2002my,
Casalbuoni:2003sa,
Casalbuoni:2004wm,Casalbuoni:2005zp,Mannarelli:2006fy,Rajagopal:2006ig,Fukushima:2006su}
(the so called LOFF phases, from condensed matter works in the
sixties~\cite{LOFF2}) appear as more promising at this time. Beside
them, gluonic phases~\cite{Gorbar:2005rx} and ground states with
spontaneous current generation~\cite{Huang:2005pv,Schafer:2005ym}
have been proposed.

One purpose of the present note will be to sharpen the diagnostic
for the chromo-magnetic pathology of the gapless phases. We shall
add a new element to such a diagnostic by finding that in addition
to the negative squared Meissner mass, which first revealed the
pathology, another element, namely a negative squared velocity of
the Goldstone related to the axial U(1)
current~\cite{Pisarski:1998nh}, is present. Such a Goldstone is not
one of the Goldstones to be eaten up by the gluons after color
gauging. We shall discuss the relation between the two elements:
Meissner masses and $U(1)_A$ Goldstone velocity. As the search for
the correct pre-asymptotic color superconductive phase cannot be
said to be concluded at this stage, a full knowledge of the possible
pathologies will presumably be useful before concluding such a
search.


We shall write down the low energy lagrangian of $\phi$, the Goldstone boson related to the spontaneous
breaking of $U(1)_A$. In the gapless regime the Goldstone particle $\phi$ is found to exhibit a
negative squared velocity. This negative velocity will be seen to be directly related to the negative
squared Meissner mass of the gluon $8^{th}$.


The relation with the Meissner chromo-magnetic instability leads to
the hope that such Goldstone pathology does not apply to the
crystalline phases, which are at this stage the favored candidates
for QCD at pre-asymptotic densities, beside gluonic phases. We shall
examine in the main part of this paper the nature of the true ground
state of two flavor superconductivity. We shall first verify the
equivalence of the one-plane-wave LOFF state with the state with
Goldstone currents, the cur-g2SC state. We shall express the ground
state in terms of the two flavor LOFF phase for one-plane-wave. We
shall then discuss the effective action for the fluctuations of the
phase of the condensates in the presence of Goldstone currents,
showing that there is no sign problem for the squared velocity. Such
a result is shown to hold also for multiple-plane-wave LOFF.
Contrary to one-plane-wave, such states may be valid candidates for
the ground state. We discuss the Meissner masses for
multiple-plane-waves, arguing that they are of positive squares.
Finally we examine the Higgs instability for multiple-plane-wave
LOFF, arguing in favor of no Higgs instability at least for small
momenta. Our conclusion is in favor of a multiple-plane-wave LOFF
for the ground state of gapless two flavor superconductors. A
comparison with the phases of condensed gluons, following the
treatment depicted in Ref.~\cite{Kiriyama:2006ui} for the case of
the single plane wave LOFF phase, is at this stage of vital
importance.

The plan of the paper is as follows: in Sec.~II we introduce the
model. In Sec.~III we derive the parameters of the small momenta
action for the Goldstone field $\phi$: the squared decay constant
and the squared velocity. In Sec.~IV we study the true ground state
of the two flavor color superconductor. We first report on the
equivalence of the one-plane-wave LOFF state with the Goldstone
current. Then we discuss the fluctuations in the cur-g2SC phase. We
write down the effective action for the Goldstone boson in the LOFF
phase and calculate the Meissner masses in the multiple-plane-wave
phases. Finally we briefly discuss the Higgs instability for such
phases. In the Appendix  we derive the screening Debye and Meissner
masses of a fictitious gauge boson. The calculation is then matched
to the low energy properties of the $\phi$.

\section{The model}\label{sec:model}
In this paper we consider two flavor superconductive quark matter
whose action is given by
\begin{equation}
S = \int d^4x \left[\bar\psi_{i\alpha}\left(i
\gamma^{\mu}\partial_\mu +
\mu_{ij}^{\alpha\beta}\gamma_0\right)\psi_{j\beta} + (L\rightarrow
R) + {\cal L}_\Delta\right]~;\label{eq:Lagr1}
\end{equation}
in the above equation $\alpha,\beta = 1,2,3$ denote color and
$i,j=1,2$ stem for flavor. The spinor $\psi$ is a left-handed Weyl
spinor. The chemical potential matrix $\mu_{ij}^{\alpha\beta}$ is
defined as
\begin{equation}
\mu_{ij}^{\alpha\beta} = \left(\mu_b \delta_{ij}-
Q_{ij}\mu_e\right)\delta_{\alpha\beta} + \left(\mu_3
T_3^{\alpha\beta} + \mu_8 T_8^{\alpha\beta}\right)\delta_{ij}~.
\label{eq:ChemPot}
\end{equation}
In Eq.~\eqref{eq:ChemPot} $\mu_b$ is one third of the baryon chemical potential; $\mu_e$ is the
electron chemical potential, coupled to quarks via the electric charge matrix $Q_{ij} =
\text{Diag}[2/3, -1/3, -1/3]$; finally $\mu_3$ and $\mu_8$ are color chemical potentials related to the
conserved charges $Q_3 = \langle\psi^\dagger T_3 \psi\rangle$ and $Q_8 = \langle\psi^\dagger T_8
\psi\rangle$. It has been shown in~\cite{Buballa:2005bv} that in the 2SC and the g2SC phases of QCD, to
which we are interested in this paper, the introduction of $\mu_3, \mu_8$ is enough  in order to
properly achieve color neutrality.

Condensation in the quark-quark channel is described by the
condensation lagrangian ${\cal L}_\Delta$ which is given by
\begin{equation}
{\cal L}_\Delta = -\frac{\Delta}{2}\psi_{i\alpha}^T C
\psi_{j\beta}\epsilon^{\alpha\beta3}\epsilon_{ij} + h.c
-(L\rightarrow R)~;\label{eq:LagrGap}
\end{equation}
it can be obtained in the mean field approximation from a local
four-fermion interaction. In writing Eq.~\eqref{eq:LagrGap} we are
assuming that in the ground state
\begin{equation}
\langle\psi_{i\alpha}^L C\psi_{j\beta}^L\rangle =
-\langle\psi_{i\alpha}^R C\psi_{j\beta}^R\rangle
\propto\Delta\epsilon^{\alpha\beta3}\epsilon_{ij}\neq0~,\label{eq:VEV}
\end{equation}
where the superscripts $L,R$ denote left-handed and right-handed quarks respectively. In the ground
state characterized by the above quark condensate one has $\mu_3 = 0$: as a matter of fact, the vacuum
expectation value~\eqref{eq:VEV} is neutral with respect to color transformations generated by $T_3$;
therefore $Q_3 = 0$ identically, and the color chemical potential associated to $T_3$ vanishes. As for
$\mu_8$, it has been shown that it is non-zero but in any case it is negligible (that is
$\mu_8\ll\mu_e$). So from now on we will put $\mu_8=0$ and assume that the only difference in chemical
potential among quarks arises from the charge chemical potential $\mu_Q = -\mu_e$.

In this paper we adopt the high density effective theory (HDET)
approximation in order to derive the quark
propagator~\cite{Nardulli:2002ma,Hong:1998tn}. This approximation is
justified since quarks live at high baryon chemical potential
$\mu_b$ and the relevant momenta for the dynamics are those near the
Fermi momentum (that in the case of massless quarks coincides with
their chemical potential). In this case, negative energy fields are
decoupled and suppressed with respect to the positive energy ones.
Thus at the leading order in the expansion $1/\mu$ one can describe
the system in terms of the positive energy fields only. In HDET one
decomposes the quark momenta as
\begin{equation}
p_0 = \ell_0~,~~~~~{\bm p} = (\mu + \ell_\parallel){\bm v} +
{\bm\ell_\perp}~,
\end{equation}
with ${\bm v}$ the Fermi velocity of the quarks, $|{\bm v}| = 1$ in
the massless case considered here. In the above equation $\mu =
(\mu_u + \mu_d)/2$. The HDET action in momentum space reads
\begin{equation}
S = \frac{\mu^2}{\pi}\int\!\frac{d {\bm
n}}{8\pi}\int\frac{d\ell_\parallel
d\ell_0}{(2\pi)^2}~\chi^\dagger\left(
\begin{array}{cc}
  V\cdot\ell + \delta\mu{\bm A} & {\bm\Delta} \\
  {\bm\Delta} & \tilde{V}\cdot\ell - \delta\mu{\bm A}
\end{array}
\right)\chi + (L\rightarrow R)\label{eq:Lagr3}
\end{equation}
where $\delta\mu = (\mu_d-\mu_u)/2$; $\chi$ is a left-handed and
positive energy velocity dependent Nambu-Gorkov field defined as
\begin{equation}
\chi = \left(\begin{array}{c}
               \Psi({\bm v}) \\
               C\Psi^*(-{\bm v})
             \end{array}
\right)~,
\end{equation}
and $\Psi
=(\psi_{ur},\psi_{ug},\psi_{dr},\psi_{dg},\psi_{ub},\psi_{db})$. In
Eq.~\eqref{eq:Lagr3} we have introduced the matrix ${\bm A}
=\text{Diag}[-1,-1,1,1,-1,1]$. Finally the gap matrix ${\bm\Delta}$
is
\begin{equation}
{\bm\Delta} = \Delta\left(\begin{array}{cccccc}
                      0 & 0 & 0 & 1 & 0 & 0 \\
                      0 & 0 & -1 & 0 & 0 & 0 \\
                      0 & -1 & 0 & 0 & 0 & 0 \\
                      1 & 0 & 0 & 0 & 0 & 0 \\
                      0 & 0 & 0 & 0 & 0 & 0 \\
                      0 & 0 & 0 & 0 & 0 & 0
                    \end{array}
\right)~.
\end{equation}

The fermion propagator $D(\ell_0,\ell_\parallel)$ for left and right
handed fields can be easily obtained from the above action once we
write it in the form
\begin{equation}
S = \frac{\mu^2}{\pi}\int\!\frac{d {\bm
n}}{8\pi}\int\frac{d\ell_\parallel d\ell_0}{(2\pi)^2}~\chi^\dagger
D(\ell_0,\ell_\parallel)^{-1}\chi + (L\rightarrow
R)~.\label{eq:prop}
\end{equation}
The poles of $D$ define the dispersion law for the quasi-particles.
They read as
\begin{eqnarray}
E_{\pm\pm} &=& \pm\delta\mu\pm\sqrt{\ell_\parallel^2 +
\Delta^2}~,~~~~~\text{red and green quarks} ~,\label{eq:dispRG}\\
E_{\pm\pm} &=& \pm\delta\mu\pm|\ell_\parallel|~,~~~~~\text{blue
quarks}~. \label{eq:dispB}
\end{eqnarray}
When $\delta\mu > \Delta$ one can introduce a blocking region (BR)
defined as
\begin{equation}
\{\ell_\parallel \ni\!{'} ~|\ell_\parallel| < \sqrt{\delta\mu^2 -
\Delta^2}\}~. \label{eq:BRdef}
\end{equation}

\section{Effective action for the Goldstone field}\label{sec:EA}
In order to derive the effective action for the Goldstone boson
$\phi$ associated to the breaking of the $U(1)_A$ symmetry in the
2SC and g2SC phases, we introduce the external field $U = \exp(i
\phi/f)$ by means of the following transformation over the fermion
fields~\cite{Eguchi:1976iz}:
\begin{equation}
\chi = \left(\begin{array}{c}
               \Psi \\
               C\Psi^*
             \end{array}
\right) \rightarrow \left(\begin{array}{c}
               U^\dagger\Psi \\
               U C\Psi^*
             \end{array}
\right)~,\label{eq:trans}
\end{equation}
so that the fermion action becomes
\begin{equation}
S = \frac{\mu^2}{\pi}\int\!\frac{d {\bm
n}}{8\pi}\int\frac{d\ell_\parallel
d\ell_0}{(2\pi)^2}~\chi^\dagger\left(
\begin{array}{cc}
  V\cdot\ell + \delta\mu{\bm A} & {\bm\Delta}e^{2i\phi/f} \\
  {\bm\Delta}e^{-2i\phi/f} & \tilde{V}\cdot\ell - \delta\mu{\bm A}
\end{array}
\right)\chi + (L\rightarrow R)~.\label{eq:LagrUU}
\end{equation}

The transformation in Eq.~\eqref{eq:trans} gives rise, at the
leading order, to a three body and a four body interaction term
among quarks and $\phi$, namely
\begin{equation}
{\cal L}_{3} =
\left(\frac{2i\phi}{f}\right)\frac{\mu^2}{\pi}\int\!\frac{d {\bm
n}}{8\pi}\int\frac{d\ell_\parallel
d\ell_0}{(2\pi)^2}~\chi^\dagger\left(
\begin{array}{cc}
  0 & {\bm\Delta} \\
  -{\bm\Delta} & 0
\end{array}
\right)\chi + (L\rightarrow R)~,\label{eq:Lagr4}
\end{equation}
\begin{equation}
{\cal L}_{4} =
-\left(\frac{2\phi^2}{f^2}\right)\frac{\mu^2}{\pi}\int\!\frac{d {\bm
n}}{8\pi}\int\frac{d\ell_\parallel
d\ell_0}{(2\pi)^2}~\chi^\dagger\left(
\begin{array}{cc}
  0 & {\bm\Delta} \\
  {\bm\Delta} & 0
\end{array}
\right)\chi + (L\rightarrow R)~.\label{eq:Lagr5}
\end{equation}
Integration over the fermion fields in the functional integral gives
rise to the effective lagrangian which, at the second order in
$\phi$, consists of two terms, namely a self-energy and a tadpole
action:
\begin{equation}
{\cal L}_{s.e.} =
\frac{i}{2}\left(\frac{2i\phi}{f}\right)^2\text{Tr}\left[D \left(
\begin{array}{cc}
  0 & {\bm\Delta} \\
  -{\bm\Delta} & 0
\end{array}
\right)D \left(
\begin{array}{cc}
  0 & {\bm\Delta} \\
  -{\bm\Delta} & 0
\end{array}
\right)\right]~,\label{eq:seLagr}
\end{equation}
\begin{equation}
{\cal L}_{tad} = -i\left(-\frac{2\phi^2}{f^2}\right)\text{Tr}\left[D
\left(
\begin{array}{cc}
  0 & {\bm\Delta} \\
  {\bm\Delta} & 0
\end{array}
\right)\right]~.\label{eq:tadLagr}
\end{equation}
Evaluation of the traces gives in momentum space
\begin{eqnarray}
i {\cal L}_{s.e.}(p_0,{\bm p}) &=&
\frac{\phi(-p)\phi(p)}{f^2}\frac{\mu^2}{\pi^3}\int\!\frac{d
{\bm n}}{8\pi}\nonumber\\
&&\times\int d\ell_\parallel
d\ell_0\left\{\frac{4\Delta^2[\Delta^2-(\ell_0+\delta\mu)(\ell_0+p_0+\delta\mu)+\ell_\parallel(\ell_\parallel+{\bm
p}\cdot{\bm n})]} {[(\ell_0+\delta\mu)^2-\ell_\parallel^2-\Delta^2]
[(\ell_0+p_0+\delta\mu)^2-(\ell_\parallel + {\bm p}\cdot{\bm n}
)^2-\Delta^2]}+\delta\mu\rightarrow-\delta\mu\right\}~\label{eq:SEmomenta}
\end{eqnarray}
and
\begin{equation}
{\cal L}_{tad} = -{\cal L}_{s.e.}(p_0 =0, {\bm
p}=0).~\label{eq:tadpoleMomenta}
\end{equation}
The last equation implies that ${\cal L}_{s.e.}(p_0 =0, {\bm p}=0) +
{\cal L}_{tad} = 0$, which is equivalent to the vanishing of the
mass of $\phi$ as required by the Goldstone theorem.

\subsection{Decay constant}
For the computation of  the decay constant $f$ it is enough to evaluate ${\cal L}_{s.e.}(p_0,{\bm p}=0)
+ {\cal L}_{tad}$. To this end, in order to properly treat the infrared divergences arising from the
gapless modes in the fermion propagator, we perform the $\ell_0$ integral at finite temperature, $\int
d\ell_0 = 2\pi i T\sum_n$ and $\ell_0 \rightarrow i \pi T (2n+1), p_0 \rightarrow i \pi T m$ in the
integrand. Once the summation over Matsubara frequencies is performed we take the limit
$T\rightarrow0$. The result is
\begin{eqnarray}
&&{\cal L}_{tad} + {\cal L}_{s.e.}(p_0,{\bm p}=0) =
\frac{\phi(-p_0)\phi(p_0)}{f^2}\frac{\mu^2}{\pi^2}
\nonumber\\
&&~~~~~~~~~~\times\int_0^{+\infty}d\ell_\parallel\left[1-\theta(\delta\mu-E(\ell_\parallel))\right]
\left[\frac{8\Delta^2}{2E(\ell_\parallel)+p_0 + i0^+} +
\frac{8\Delta^2}{2E(\ell_\parallel)-p_0-i0^+} -
\frac{8\Delta^2}{E(\ell_\parallel)}\right]~.\label{eq:sum1}
\end{eqnarray}
In the above relation $E(x) =\sqrt{x^2 + \Delta^2}$. In the
analytical continuation from imaginary to real boson energy we add a
small positive imaginary part to $p_0$. As usual, this gives rise to
an imaginary part of the polarization tensor for the boson, which is
related to its decay rate. The imaginary part develops when $p_0 >
2\Delta$, the sum of the rest energies of the two quasi-particles.
This is equal to what happens in the gapped 2SC phase, when
$\delta\mu=0$. The integral over quark momentum can be performed
analytically for each value of $p_0$; at the order $p_0^2/\Delta^2$
one gets
\begin{equation}
{\cal L}_{tad} + {\cal L}_{s.e.}(p_0,{\bm p}=0) =
\frac{1}{2f^2}(p_0\phi)(p_0
\phi)\frac{4\mu^2}{\pi^2}\left(1-\theta(\delta\mu-\Delta)
\frac{\sqrt{\delta\mu^2-\Delta^2}}{\delta\mu}\right)~.\label{eq:Fsq}
\end{equation}
Imposing the canonical normalization of the field $\phi$ in
Eq.~\eqref{eq:Fsq} fixes the value of $f$, namely
\begin{equation}
f^2 = \frac{4\mu^2}{\pi^2}\left(1-\theta(\delta\mu-\Delta)
\frac{\sqrt{\delta\mu^2-\Delta^2}}{\delta\mu}\right)~.\label{eq:Fsq2}
\end{equation}
When $\delta\mu=0$ we obtain the well known result of the 2SC phase,
$f^2_{2SC} = 4\mu^2/\pi^2$. In the limit $\delta\mu \gg \Delta$ one
has
\begin{equation}
f^2 \approx
\frac{2\mu^2}{\pi^2}\frac{\Delta^2}{\delta\mu^2}~.\label{eq:fSqAsym}
\end{equation}



\subsection{Squared velocity}
Next we turn to the computation of the squared velocity of $\phi$.
To this end it is enough to consider ${\cal L}_{s.e.}(0,{\bm p}) +
{\cal L}_{tad}$. The integral over quark energy is performed as
above at finite temperature, keeping the limit $T\rightarrow0$ after
the summation on the Matsubara frequencies is done. We find
\begin{equation}
{\cal L}_{s.e.}(0,{\bm p}) + {\cal L}_{tad} =
\frac{\phi(-p)\phi(p)}{f^2}\frac{\mu^2}{\pi^2}\int\!\frac{d {\bm
n}}{8\pi}\int_{-\infty}^{+\infty}d\ell_\parallel\left[F(\ell_\parallel,{\bm
p} \cdot {\bm n}) -  F(\ell_\parallel,0) \right]~\label{eq:FFF}
\end{equation}
with
\begin{equation}
F(x,y) = \frac{4\Delta^2}{2x+y}\frac{2x}{\sqrt{x^2+\Delta^2}}
 \left(1-\theta(\delta\mu-\sqrt{x^2+\Delta^2})\right) + \frac{4\Delta^2}{2x+y}\frac{2(x+y)}{\sqrt{(x+y)^2+\Delta^2}}
 \left(1-\theta(\delta\mu-\sqrt{(x+y)^2+\Delta^2})\right)~.
\end{equation}
Once the integral over the quark momentum is performed (using the
$\theta-$functions) we expand at $O(p^2/\Delta^2)$; we find
\begin{eqnarray}
{\cal L}_{s.e.}(0,{\bm p}) + {\cal L}_{tad} &=&
-\frac{1}{2f^2}(p_i\phi)(p_i\phi) \frac{4\mu^2}{3\pi^2}
\left(1-\theta(\delta\mu-\Delta)\frac{\delta\mu}{\sqrt{\delta\mu^2-\Delta^2}}\right)
\nonumber\\
&\equiv& - \frac{v^2}{2}(p_i\phi)(p_i\phi)~,
\end{eqnarray}
with the squared velocity defined as
\begin{equation}
v^2 = \frac{4\mu^2}{3\pi^2 f^2}
\left(1-\theta(\delta\mu-\Delta)\frac{\delta\mu}{\sqrt{\delta\mu^2-\Delta^2}}\right)
\label{eq:vSq1}
\end{equation}
and $f^2$ given by Eq.~\eqref{eq:Fsq2}. For $\delta\mu=0$ we recover
the well known result $v^2_{2SC} = 1/3$. With these results at hand
we can write the effective action of $\phi$ at small momenta as
\begin{equation}
S_{eff}[\phi] = \int d^4x\frac{1}{2}\left((\partial_0\phi)^2 -
v^2(\bm{\nabla}\phi\bm{\nabla}\phi)\right)~.\label{eq:EffActEya}
\end{equation}
Eq.~\eqref{eq:EffActEya}, with $f^2$ and $v^2$ given respectively in Eqs.~\eqref{eq:Fsq2}
and~\eqref{eq:vSq1}, is one of the results of this paper.

Notice that in the gapless phase we find a negative squared velocity
of $\phi$. This instability in the Goldstone sector is directly
related to the chromo-magnetic instability of the gapless 2SC phase.
As a matter of fact, we see that we can write the relation
\begin{equation}
v^2 \propto m_{M,8}^2
\end{equation}
where $m_{M,8}^2$ is the squared Meissner mass of the $8^{th}$ gluon
calculated in~\cite{Huang:2004bg}
\begin{equation}
m_{M,8}^2 = \frac{4\alpha_s\mu^2}{9\pi}
\left(1-\theta(\delta\mu-\Delta)\frac{\delta\mu}{\sqrt{\delta\mu^2-\Delta^2}}\right)~.
\end{equation}

As far as the gapped phases, the relation between the squared
velocity of the Goldstone mode and the squared Meissner mass of a
gluon (or, more generally, of a spin $1$ gauge field) is not new,
see~\cite{Son:1999cm,Beane:2000ms,Schafer:2001za,Manuel:2000wm} for
2SC and CFL. It is interesting to notice that the $v^2$ of $\phi$ is
negative only in the gapless regime $\delta\mu
> \Delta$, while the 2SC phase presents the chromo-magnetic
instability also in the gapped region $\Delta/\sqrt{2} < \delta\mu < \Delta$: in this interval one has
$m_{M,8}^2 > 0$ but $m_{M,a}^2 < 0$, with $a = 4,\cdots,7$. As shown  in~\cite{Gorbar:2005rx} the
instability in the gapped regime can be cured by a gluonic condensate. The method of
Ref.~\cite{Gorbar:2005rx} however can be applied only in the region $\delta\mu\approx\Delta/\sqrt{2}$,
where the gluon condensates are expected to be small, because it relies on the expansion of the
expectation values of some of the gluonic fields around zero, see~\cite{Gorbar:2005rx} for more
details.
In the gapless regime $\delta\mu
> \Delta$ the instability could be cured either by a gluon condensate or by net baryon and/or meson
currents, see~\cite{Huang:2005pv} for a wide discussion. We expect
that the removal of the chromo-magnetic instability for the $8^{th}$
gluon is enough for the removal of the Goldstone mode instability.
Finally, we notice that $v^2 < 0$ for the Goldstone mode $\phi$ is
similar to the instability found in Ref.~\cite{Gorbar:2006up} for
the $8^{th}$ gluon.


\section{The genuine ground state}
In the previous section we have computed the low energy parameters of the Goldstone mode $\phi$ both in
the 2SC and in the g2SC phase. In particular, we have found a negative squared velocity, a signal of an
instability of the ground state. Therefore, the g2SC can not be the true vacuum of the model and one
has to look for other solutions. In this section we introduce an ansatz for the true ground state,
which imitates the LOFF phase of condensed matter. This topic is not new: both in the two flavor and in
the three flavor gapless phases of QCD it has been extensively studied~\cite{Huang:2005pv}. However, it
has never been related to the negative squared velocity of one (or more) Goldstone bosons. We cover
this topic in this section.

\subsection{The one plane wave LOFF state and its equivalence with the Goldstone current}
The starting point is the observation that, because of the wrong sign in the spatial part of the
kinetic term in Eq.~\eqref{eq:EffActEya}, the expectation value for ${\bm\nabla}\phi$ (VEV) may be
expected to be different from zero. In order to determine the true ground state we decompose the $\phi$
around a mean field part, and study the effective action for the fluctuations around this mean field.
Moreover, since ${\bm\nabla}\phi$ is a vector, the VEV breaks spontaneously the rotational symmetry.
The simplest way to introduce the fluctuations is writing $\phi$ as
\begin{equation}
\phi(x) = {\bm\Phi}\cdot{\bm x} + h(x)~, \label{eq:fluct1}
\end{equation}
where ${\bm\Phi}$ is a constant and homogeneous vector and $h$ is
the fluctuation field; in this way one has
\begin{equation}
{\bm\nabla}\phi = {\bm\Phi} + {\bm\nabla}h~;
\end{equation}
assuming $\langle{\bm\nabla}h\rangle = 0$ we are left with $\langle{\bm\nabla}\phi\rangle = {\bm\Phi}$.
Then, choosing ${\bm\Phi}$ appropriately, one has to show that the lagrangian of the fluctuations
$h(x)$ does not suffer of the sign problem in the spatial part. Since ${\bm\Phi} \neq 0$ corresponds to
a Goldstone current, we call the ground state cur-g2SC phase.

Introducing Eq.~\eqref{eq:fluct1} into Eq.~\eqref{eq:LagrUU} one notices that the action resembles that
for the two flavor LOFF phase in the one plane wave (1PW) structure,
\begin{equation}
S = \int d^4 x \int\!\frac{d {\bm n}}{8\pi}~\chi^\dagger\left(
\begin{array}{cc}
  iV\cdot\partial + \delta\mu{\bm A} & {\bm\Delta}\exp\left\{\displaystyle\frac{2i({\bm\Phi}\cdot{\bm x} + h)}{f}\right\} \\
  {\bm\Delta}\exp\left\{\displaystyle\frac{-2i({\bm\Phi}\cdot{\bm x} + h)}{f}\right\}  & i\tilde{V}\cdot\partial - \delta\mu{\bm A}
\end{array}
\right)\chi + (L\rightarrow R)~,\label{eq:LagrUU2}
\end{equation}
with ${\bm\Phi}/f \equiv {\bm q}$ playing the role of the wave vector and the field $h$ being the
phonon~\cite{Casalbuoni:2002pa} as well as the Goldstone boson related to the $U(1)_A$. Since the
fermion dispersion laws in 1PW and in cur-g2SC are the same, the thermodynamic behavior  of the two
phases is indistinguishable and in what follows we can refer both as the 1PW and as the cur-g2SC.
However we stress that the cur-g2SC phase is built with zero momentum Cooper pairs, and the wave vector
${\bm q}$ in this context is related to the non-vanishing Goldstone current in the ground state. On the
other hand, the 1PW ground state is built with Cooper pairs with a total momentum equal to $2{\bm q}$;
the Goldstone current in this case is vanishing, as we prove in the next section.

The 1PW state can be  analyzed exactly, and the effective action for
the fluctuation $h$ can be determined without approximations: it is
enough to shift the quark momenta by an amount $\pm{\bm q}$, where
the upper and lower signs stem respectively for $u$ and $d$ quarks.
In this way the ${\bm x}$ dependence in the gap term is ruled out in
place of a shift $\delta\mu \rightarrow \delta\mu - {\bm q} \cdot
{\bm n}$ in the quark chemical potentials; but in this case the
expressions involved in the calculations of the loop integrals are
more complicated than the homogeneous ones. Therefore we prefer to
treat this problem by a Ginzburg-Landau (GL) expansion of the
fermion propagator in $\Delta/\delta\mu$. In expanding {\em a la}
Ginzburg-Landau we pay the price of loosing exact expressions,
obtaining equations valid only to a fixed order in
$\Delta/\delta\mu$, but the formula are easy to handle and to
generalize to the case of more complicated crystal structures (see
below).

In the GL approximation one formally writes the quark propagator,
which can be read from Eq.~\eqref{eq:Lagr3}, as
\begin{equation}
D = \sum_{n=1}^\infty (-1)^n\left[D_0 \left(\begin{array}{cc}
                                              0 & {\bm\Delta} \\
                                              {\bm\Delta} & 0
                                            \end{array}
\right)\right]^n D_0~,
\end{equation}
where $D_0$ is the propagator at $\Delta = 0$. The zero temperature
thermodynamic potential for the quarks in this approximation
reads~\cite{Bowers:2002xr}
\begin{equation}
\Omega = \Omega_0 + \frac{\alpha}{2}\Delta^2 +
\frac{\beta}{4}\Delta^4 + \frac{\gamma}{6}\Delta^6 + {\cal
O}(\Delta^8)~, \label{eq:OmegaOOO}
\end{equation}
where the coefficients are given by
\begin{equation}
\alpha = - \frac{4\mu^2}{\pi^2}\left(1-\frac{\delta\mu}{2 |{\bm q}|}
 \log\left|\frac{|{\bm q}| + \delta\mu}{|{\bm q}| - \delta\mu}\right|
 + \frac{1}{2}\log\frac{\Delta_0^2}{4(|{\bm q}|^2 - \delta\mu^2)}
 \right)~,
\end{equation}
\begin{equation}
\beta = \frac{\mu^2}{\pi^2}\frac{1}{|{\bm q}|^2 -
\delta\mu^2}~,~~~~~\gamma =~\frac{\mu^2}{8\pi^2}\frac{|{\bm q}|^2 +
\delta\mu^2}{(|{\bm q}|^2-\delta\mu^2)^3}~,
\end{equation}
and $\Omega_0$ is the free gas contribution. The physical value of
${\bm q}$ is obtained as usual by minimization of the thermodynamic
potential. At the leading order in $\Delta/\delta\mu$ one has
\begin{equation}
\left.\frac{\partial\Omega}{\partial |{\bm q}|}\right|_{|{\bm q}| =
Q} = 0 \Leftrightarrow  1-\frac{\delta\mu}{2 Q}
 \log\left|\frac{Q + \delta\mu}{Q - \delta\mu}\right|
 = 0~,  \label{eq:MinOmegaQ}
\end{equation}
which gives the result $Q \simeq 1.2 \delta\mu$ well known in the
LOFF literature. Since ${\bm \Phi}^2 =f^2 {\bm q}^2$, once we know
$f^2$ we are able to evaluate the Goldstone current in the cur-g2SC
state.

\subsection{Effective action of the fluctuation in the cur-g2SC state}
Next we turn to the effective action for the fluctuation field $h$.
Evaluating the traces in Eqs.~\eqref{eq:seLagr}
and~\eqref{eq:tadLagr} at the leading order in $\Delta/\delta\mu$
one is left with the expression
\begin{equation}
{\cal L}_{s.e.}(p) + {\cal L}_{tad} = -i\frac{2\Delta^2}{f^2} h(-p)
 \left[ {\cal J}(p) - {\cal J}(0) \right]h(p)~, \label{eq:hhh}
\end{equation}
where $p = (p_0,{\bm p})$ and the loop integral ${\cal J}(p)$ is
defined as
\begin{eqnarray}
{\cal J}(p) &=& -2\int\!\frac{d{\bm n}}{4\pi} \int\!\frac{d^4
\ell}{(2\pi)^4} \left[\frac{1}{(\tilde
V\cdot\ell - \delta\mu + {\bm q}\cdot{\bm n})[V\cdot(\ell + p) - \delta\mu + {\bm q}\cdot{\bm n}] } \right. \nonumber \\
&& ~~~~~~~~~~~~~~~~~~~~~~~ + \left.\frac{1}{( V\cdot\ell - \delta\mu
+ {\bm q}\cdot{\bm n})[\tilde V\cdot(\ell + p)  - \delta\mu + {\bm
q}\cdot{\bm n}] } \right] + \delta\mu \rightarrow
-\delta\mu~.\label{eq:PaI}
\end{eqnarray}

From now on the calculation is similar to the one presented in great
detail in~\cite{Mannarelli:2007bs} for the displacement fields in
the three flavor LOFF phase of QCD; therefore in this paper we
simply show the main steps of the calculation, referring the
interested reader to Ref.~\cite{Mannarelli:2007bs} for further
details. For small external momenta $p$ one has
\begin{equation}
{\cal J}(p) - {\cal J}(0) = 2\int\!\frac{d{\bm
n}}{4\pi}\int\!\frac{d^4 \ell}{(2\pi)^4}\frac{\tilde V\cdot p~V\cdot
p}{(\tilde V\cdot\ell - \delta\mu + {\bm q}\cdot{\bm n})^2
(V\cdot\ell - \delta\mu + {\bm q}\cdot{\bm n})^2} + \delta\mu
\rightarrow -\delta\mu~.
\end{equation}
The computation of the loop integral is done in the usual way by
Wick rotating to imaginary energies $\ell_0 \rightarrow i\ell_4$;
since the integral is convergent one can send the ultraviolet cutoff
on $\ell_\parallel$ to infinity, and perform the integral over
$\ell_\parallel$ by residues, followed by integration over $\ell_4$.
This is the same procedure used for the calculation of the
coefficients $\beta$, $\gamma$ in the GL effective
potential~\cite{Bowers:2002xr}. We find
\begin{equation}
{\cal J}(p) - {\cal J}(0) = -i\frac{\mu^2}{4\pi^2}\Re
e\int\!\frac{d{\bm n}}{4\pi}\frac{\tilde V\cdot p~ V\cdot
p}{(\delta\mu - {\bm q}\cdot {\bm n} + i0^+)^2} + \delta\mu
\rightarrow -\delta\mu~. \label{eq:Pfull}
\end{equation}

From Eqs.~\eqref{eq:hhh} and~\eqref{eq:Pfull} one can easily read
the low energy parameters of the effective lagrangian for the
fluctuation field in the cur-g2SC phase, namely
\begin{equation}
f^2 = - \frac{\Delta^2 \mu^2}{\pi^2}\Re e\int\!\frac{d{\bm
n}}{4\pi}\frac{1}{(\delta\mu - {\bm q}\cdot {\bm n} + i0^+)^2}
+\left( \delta\mu \rightarrow -\delta\mu\right) ~=~
\frac{2\mu^2}{\pi^2}\frac{\Delta^2}{Q^2 - \delta\mu^2}~,
\label{eq:fSqLOFF}
\end{equation}
\begin{equation}
v^2_x = v^2_y = - \frac{\Delta^2 \mu^2}{f^2\pi^2}\Re
e\int\!\frac{d{\bm n}}{4\pi}\frac{n_x^2}{(\delta\mu - {\bm q}\cdot
{\bm n} + i0^+)^2} + \left( \delta\mu \rightarrow -\delta\mu \right)
~=~ 0~,\label{eq:VsqxLOFF}
\end{equation}
\begin{equation}
v^2_z =  - \frac{\Delta^2 \mu^2}{f^2\pi^2}\Re e\int\!\frac{d{\bm
n}}{4\pi}\frac{n_z^2}{(\delta\mu - {\bm q}\cdot {\bm n} + i0^+)^2} +
\left( \delta\mu \rightarrow -\delta\mu \right) ~=~
\frac{2\mu^2}{f^2 \pi^2}\frac{\Delta^2}{Q^2 -
\delta\mu^2}~.\label{eq:VsqzLOFF}
\end{equation}
Finally, in configuration space, the lagrangian of the fluctuation
reads
\begin{equation}
{\cal L}[h] = \frac{1}{2}\left((\partial_0 h)^2 - {\bm v}\cdot({\bm
\nabla}h)~ {\bm v}\cdot({\bm \nabla}h)  \right)~.
\label{eq:effactkkk}
\end{equation}

By means of Eq.~\eqref{eq:fSqLOFF} we determine the value of the
Goldstone current which minimizes the effective potential,
\begin{equation}
|\langle {\bm\nabla}\phi\rangle|^2 = |{\bm\Phi}|^2 =
\frac{2\mu^2}{\pi^2}\Delta^2\frac{Q^2}{Q^2 - \delta\mu^2} \approx
0.66 \mu^2 \Delta^2~,
\end{equation}
where we have used the relation $Q \simeq 1.2 \delta\mu$. Moreover
Eqs.~\eqref{eq:VsqxLOFF} and~\eqref{eq:VsqzLOFF} show that the
squared velocity of $h$ is {\em positive} along the direction of
${\bm\Phi}$ and is {\em zero} in the plane orthogonal to
${\bm\Phi}$. As can be easily shown by direct calculation, the zero
value of the orthogonal velocity is due to the proportionality
relation between the integral in Eq.~\eqref{eq:VsqxLOFF} and the
derivative $\partial\Omega/\partial |{\bm q}|$ evaluated at the
minimum, see Eq.~\eqref{eq:MinOmegaQ}. Therefore, the fluctuation
$h$ does not suffer the sign problem of the squared velocity, as
anticipated.

\subsection{Effective action of the Goldstone boson in the LOFF phase}
In the previous section we have shown that assuming the existence of a Goldstone current in the ground
state of the gapless 2SC quark matter, and expanding the Goldstone field around the VEV as in
Eq.~\eqref{eq:fluct1}, the lagrangian of the fluctuation does not suffer the sign problem of the
squared velocity. The cur-g2SC phase resembles the one plane wave LOFF state in the sense that the
breaking of the translational and rotational symmetries due to ${\bm\Phi} \neq 0$ is not
distinguishable from the breaking due to a net momentum of the Cooper pair, as can be seen by
Eq.~\eqref{eq:LagrUU2}. Therefore, it is obvious that the effective lagrangian of the Goldstone of
$U(1)_A$ in the 1PW LOFF phase is equal to the lagrangian of the fluctuation found in the previous
section. In the 1PW phase the Goldstone does not suffer of the wrong sign problem, and the free energy
of the ground state with Goldstone current is equal to the free energy of the 1PW phase.

In the 1PW phase the gap parameter has an explicit spatial
dependence of the form
\begin{equation}
\Delta({\bm r})_{1PW} = \Delta~e^{2i{\bm q}\cdot{\bm r}}~,
\end{equation}
with ${\bm q}$ being the wave vector (we have dropped for simplicity
the color, flavor and Dirac indices), equal to one half of the total
momentum of the Cooper pairs. Because of the anisotropy, only a
fraction of the Fermi surfaces of the quarks are available for the
pairing: this results in $\Delta$ (and consequently, $\Omega$)
smaller than the one of the BCS phase. However the free energy of
one plane wave LOFF state can be lowered by summing up $P$ plane
waves (PPW)~\cite{Bowers:2002xr},
\begin{equation}
\Delta({\bm r})_{PPW} = \Delta~\sum_{a=1}^{P}e^{2i{\bm q}_a\cdot{\bm
r}}~; \label{eq:GapLPFFcc}
\end{equation}
the resulting phase is known as a crystalline superconductor, as the behavior of the gap in the
configuration space resembles that of a crystal lattice.

Since the free energy of the LOFF phase with order parameter given
by Eq.~\eqref{eq:GapLPFFcc} is lower than the single plane wave one,
it follows that the 1PW can not be the ground state of a crystalline
superconductor. Stated in other words, since the cur-g2SC and the
1PW phases are not distinguishable, the free energy of the cur-g2SC
phase is higher than the free energy of a PPW crystalline state.
Therefore it seems that the cur-g2SC can not be the ground state of
two flavor quark matter, as it can be easily replaced by crystalline
phases. This situation is quite different from the three flavor
case. As a matter of fact, it has been shown
in~\cite{Schafer:2005ym} that a Goldstone current there exists in
the ground state, near the onset CFL$\rightarrow$gCFL. The curCFL
phase considered in~\cite{Schafer:2005ym} is not likely to be
replaced by a multiple plane wave LOFF state near the onset since
the free energy of the latter, as evaluated
in~\cite{Rajagopal:2006ig}, is higher than the gCFL one and
therefore still higher than the energy of the curCFL phase.

Since in the crystalline LOFF phases there exists the Goldstone
related to the breaking of $U(1)_A$, the calculation of its
effective action in the PPW state becomes of vital importance.
Nevertheless, we learn from the 1PW an important lesson: the
effective lagrangian, at least in this simple case, does not suffer
the sign problem. We wish to verify that the same property is valid
for a multiple plane wave crystalline superconductor, which is a
better candidate for the ground state.

In the PPW phase the inverse quark propagator can not be inverted,
so one is forced to make some approximation in order to write a
propagator. As anticipated in the previous section we employ the GL
expansion. In this case, at the leading order in $\Delta/\delta\mu$,
the value of $|{\bm q}|$ which minimizes the thermodynamic potential
is again given by Eq.~\eqref{eq:MinOmegaQ}, in which now $Q$
represents the equilibrium value of the total momentum of the
pairs~\cite{Bowers:2002xr} (in the cur-g2SC context it is related
instead to the VEV of the Goldstone current ).

The calculation of the effective action can be done following the
same steps outlined in the previous section, replacing the
fluctuation $h$ with the Goldstone $\phi$. As shown explicitly
in~\cite{Mannarelli:2007bs}, in the PPW state Eq.~\eqref{eq:Pfull}
has to be replaced with
\begin{equation}
{\cal J}(p) - {\cal J}(0) = -i\frac{\mu^2}{4\pi^2}\Re
e\sum_{a=1}^{P}\int\!\frac{d{\bm n}}{4\pi}\frac{\tilde V\cdot p~
V\cdot p}{(\delta\mu - {\bm q}_a\cdot {\bm n} + i0^+)^2} + \delta\mu
\rightarrow -\delta\mu~. \label{eq:PfullPPW}
\end{equation}
In the above equation the interference terms that mix different
${\bm q}_a$ do not appear: this is due to momentum conservation in
the loop integral, which forces the wave vector of the two gap
insertions to be equals. From Eq.~\eqref{eq:PfullPPW} one has for
the squared decay constant
\begin{equation}
f^2 = \frac{2\mu^2}{\pi^2}\frac{P \Delta^2}{Q^2 - \delta\mu^2}~,
\label{eq:fSqLOFFpp}
\end{equation}
with $Q \simeq 1.2 \delta\mu$. For what concerns the squared
velocity it is convenient to introduce the matrix
\begin{equation}
{\cal V}_{ij}=  - \frac{\Delta^2 \mu^2}{f^2\pi^2}\Re
e\sum_{a=1}^{P}\int\!\frac{d{\bm n}}{4\pi}\frac{n_i n_j}{(\delta\mu
- {\bm q}_a\cdot {\bm n} + i0^+)^2} + \left( \delta\mu \rightarrow
-\delta\mu \right) ~;\label{eq:VsqPPW}
\end{equation}
the elements of the matrix ${\cal V}$ are easily evaluated: following the strategy depicted
in~\cite{Casalbuoni:2002my}, for each angular integral in the sum one rotates the ${\bm q}_a$ along the
$z-$axes by means of the orthogonal matrix $R^a$. In this way one is left with the expression
\begin{equation}
{\cal V}_{ij}=  \frac{1}{P} \sum_{a=1}^{P} R^a_{3i} R^a_{3j}
~,\label{eq:VsqPPW2}
\end{equation}
where we have used Eq.~\eqref{eq:fSqLOFFpp} and the fact that at the
minimum only the longitudinal integrals are not vanishing, see
Eqs.~\eqref{eq:VsqxLOFF} and~\eqref{eq:VsqzLOFF}. With this
definition the effective action for the Goldstone reads, in momentum
space,
\begin{equation}
{\cal L}[\phi] = \frac{1}{2}\phi(-p)\left(p_0^2  - {\cal
V}_{ij}~{\bm p}_i {\bm p}_j  \right) \phi(p)~. \label{eq:effactPPW}
\end{equation}

The dispersion law $E({\bm p})$ of the Goldstone is the solution of
the equation
\begin{equation}
E({\bm p})^2 - {\cal V}_{ij}~{\bm p}_i {\bm p}_j =0~.
\end{equation}
As a consequence, in order to show that the lagrangian does not
suffer the sign problem, it is sufficient to show that the matrix
${\cal V}$ is semi-definite (or definite) positive, that is for each
${\bm p} \neq 0$ it satisfies the condition ${\cal V}_{ij}~{\bm p}_i
{\bm p}_j \geqslant 0$ (or ${\cal V}_{ij}~{\bm p}_i {\bm p}_j > 0$).
This is an easy task: as a matter of fact, from the very definition
of the rotation matrices $R^a$ it follows that one can write ${\cal
V}_{ij} = \sum_a \hat{q}^a_i \hat{q}^a_j /P$, with $\hat{q}^q_i$ a
unit vector parallel to ${\bm q}^a_i$; since the tensor $\hat{q}^a_i
\hat{q}^a_j$ is semi-definite positive, as it has one eigenvalue 1
and two eigenvalues 0, then ${\cal V}_{ij}$ is semi-definite or
definite positive, depending on the particular crystalline structure
considered. We show this in three interesting cases.

In the first case we consider a crystal with three mutually
orthogonal wave vectors (3PW). Each of these vectors can be chosen
along one of the three axes,
\begin{equation}
{\bm q}_1 = Q(1,0,0)~,~~~{\bm q}_2 = Q(0,1,0)~,~~~{\bm q}_3 =
Q(0,0,1)~.
\end{equation}
The second structure we consider is the body centered cube crystal
(BCC), defined by the following wave vectors:
\begin{eqnarray}
&&{\bm q}_1 = Q(1,0,0)~,~~~{\bm q}_2 = Q(0,1,0)~,~~~{\bm q}_3 =
Q(0,0,1)~,\nonumber \\
&&{\bm q}_4 = -{\bm q}_1~,~~~~~~~~~~{\bm q}_5 = -{\bm
q}_2~,~~~~~~~~~~{\bm q}_6 =-{\bm q}_3~.
\end{eqnarray}
Finally we consider the case of the face centered cube structure
(FCC), whose wave vectors are given by
\begin{eqnarray}
&&{\bm q}_1=\frac{Q}{\sqrt 3}(+1,+1,+1),~~~{\bm q}_2=\frac{Q}{\sqrt
3}(+1,-1,+1),~\cr &&{\bm q}_3=\frac{Q}{\sqrt 3}(-1,-1,+1),~~~{\bm
q}_4=\frac{Q}{\sqrt 3}(-1,+1,+1),\cr &&{\bm q}_5=\frac{Q}{\sqrt
3}(+1+,1,-1),~~~{\bm q}_6=\frac{Q}{\sqrt 3}(+1,-1,-1),\cr &&{\bm
q}_7=\frac{Q}{\sqrt 3}(-1,-1,-1),~~~{\bm q}_8=\frac{Q}{\sqrt
3}(-1,+1,-1)~ .
\end{eqnarray}
The rotation matrices $R^a$ are trivial for both the case of the 3PW and the  BCC crystals. As for the
FCC they can be found in the appendix B of~\cite{Casalbuoni:2002my}. For all of the crystals we find
${\cal V}_{ij} = \delta_{ij}/3$, which implies that the dispersion law of the Goldstone is
\begin{equation}
E^2({\bm p}) = \frac{1}{3}{\bm p}^2~,
\end{equation}
and the squared velocity does not suffer the wrong sign problem.

From the results of~\cite{Bowers:2002xr,Casalbuoni:2004wm} we can
infer that the BCC and the FCC may exist in a large window of the
ratio $\delta\mu/\Delta_0$, both on the left and on the right of the
Clogstone limit $\Delta_0/\sqrt{2}$. Since our result shows that in
these crystalline phases the Goldstone does not suffer the
instability towards the formation of a current, we argue that such
phases could be the genuine ground state of two flavor quark matter
at high density. A more accurate study is needed at this point:
indeed, it would be interesting to build electrical and color
neutral crystalline phases in two flavor quark matter, and to
compare their free energy with the gluonic phases and/or the gapped
2SC.

\subsection{Meissner masses of the gluons in the PPW LOFF phase}
In this subsection we compute the Meissner masses of the gluons in
the multiple plane wave state. The calculation is done in the
Ginzburg-Landau approximation at the second order in
$\Delta/\delta\mu$. The Meissner tensor can be defined
as~\cite{Nardulli:2002ma}
\begin{equation}
\left(M_{ab}^{ij}\right)^2 = -\Pi_{ab}^{ij}(p_0=0,{\bm
p}\rightarrow0)~,
\end{equation}
where the polarization tensor in the HDET approach is given by the
sum of two contributions, $\Pi_{ab}^{\mu\nu}(p) = {\cal
S}^{\mu\nu}_{ab}(p) + {\cal T}^{\mu\nu}_{ab}$, where ${\cal
S}^{\mu\nu}_{ab}(p)$ is a self-energy diagram,
\begin{equation}
{\cal S}^{\mu\nu}_{ab}(p) = i\frac{2\mu^2 }{\pi}\int\!\frac{d {\bm
n}}{8\pi}\int \frac{d\ell_\parallel d\ell_0}{(2\pi)^2} \text{Tr}
\left[D(\ell+p)\left(\begin{array}{cc}
                                 -V^\mu T_a & 0 \\
                                 0 & \tilde{V}^\mu T_a^*
                               \end{array}
\right)D(\ell)\left(\begin{array}{cc}
                                 -V^\nu T_b & 0 \\
                                 0 & \tilde{V}^\nu T_b^*
                               \end{array}
\right)\right] ~,
\end{equation}
and ${\cal T}^{\mu\nu}_{ab}$ is a tadpole diagram,
\begin{equation}
{\cal T}^{\mu\nu}_{ab} = -i\frac{2\times2}{\pi}\int\!\frac{d {\bm
n}}{8\pi}\int \frac{d\ell_\parallel d\ell_0}{(2\pi)^2}
\text{Tr}\left[D(\ell) \left(
\begin{array}{cc}
  \frac{(\mu + \ell_\parallel)^2}{2\mu+\tilde{V}\cdot\ell}T_a T_b  & 0 \\
  0 & \frac{(\mu + \ell_\parallel)^2}{2\mu+V\cdot\ell}T_a^* T_b^*
\end{array}
\right)\right]P_{\mu\nu}~. \label{eq:tadMa}
\end{equation}
In the above equations $D(\ell)$ is the fermion propagator, which will be evaluated in the following in
the GL approximation. $T_a$ is the $SU(3)$ color generator in the basis $\Psi =
(\psi_{ur},\psi_{ug},\psi_{dr},\psi_{dg},\psi_{ub},\psi_{db})$, with the normalization $\text{Tr}[T_a
T_b] = \delta_{ab}/2$. Finally the projector $P_{\mu\nu}$ is defined as
\begin{equation}
P^{\mu\nu} = g^{\mu\nu} - \frac{1}{2}\left[V^\mu V^\nu +
\tilde{V}^\mu \tilde{V}^\nu\right]~.
\end{equation}
The trace in the diagrams have to be evaluated in NG as well as in
color-flavor indices.

Technically, the evaluation of the loop integrals for the Meissner masses is similar to that involved
in the calculation of the decay constant and the velocity of the Goldstone boson: for all of the gluons
one is left with two kinds of integrals, namely
\begin{eqnarray}
I_{ij} &=& \sum_{a=1}^{P} \int\!\frac{d{\bm n}}{4\pi} \int
\frac{d\ell_\parallel d\ell_0}{(2\pi)^2} \frac{n_i n_j}{(V\cdot\ell
- \delta\mu)^3(\tilde V\cdot\ell - \delta\mu + 2{\bm q}_a\cdot{\bm
n})} + \delta\mu \rightarrow -\delta\mu ~ \nonumber\\
&=& -\frac{i}{16\pi}\Re e\sum_{a=1}^{P}\int\!\frac{d{\bm
n}}{4\pi}\frac{n_i n_j}{(\delta\mu - {\bm q}_a\cdot {\bm n} +
i0^+)^2} + \left( \delta\mu \rightarrow -\delta\mu \right)~,
\end{eqnarray}
related to a diagram with two gap insertions on a fermion branch and
zero insertions on the other branch, and
\begin{eqnarray}
J_{ij} &=& \sum_{a=1}^{P} \int\!\frac{d{\bm n}}{4\pi} \int
\frac{d\ell_\parallel d\ell_0}{(2\pi)^2} \frac{n_i n_j}{(V\cdot\ell
- \delta\mu)^2(\tilde V\cdot\ell - \delta\mu + 2{\bm q}_a\cdot{\bm
n})^2} + \delta\mu \rightarrow -\delta\mu = 2 I_{ij}~,
\end{eqnarray}
related to a diagram with one gap insertions on a fermion branch and
one gap insertion on the other branch. Because of the
proportionality relation between $I_{ij}$ and $J_{ij}$, one can
express the Meissner tensor for each crystal structure only in terms
of $I_{ij}$. At the leading order in $\Delta/\delta\mu$ we find
\begin{eqnarray}
&&\left(M_{11}^{ij}\right)^2  = \left(M_{22}^{ij}\right)^2
=\left(M_{33}^{ij}\right)^2  =0 ~,\\
&&\left(M_{aa}^{ij}\right)^2  = -i\frac{\Delta^2\mu^2}{\pi} I_{ij}~,
~~~~~a=4,\dots,7~,\label{eq:4loff}\\
&&\left(M_{88}^{ij}\right)^2 = \frac{4}{3}\left(M_{44}^{ij}\right)^2
~.\label{eq:8loff}
\end{eqnarray}

The above results are in agreement with Eqs.~(93),~(94) of Ref.~\cite{Giannakis:2005vw}, where they
have been deduced for the first time in the single plane wave LOFF state and have been generalized here
to the case of a generic crystalline structure (for simplicity here we have neglected the mixing of the
$8^{th}$ gluon with the photon. If one keeps into account this effect, only the overall factor in
Eq.~\eqref{eq:8loff} changes, as shown in~\cite{Giannakis:2005vw}). Looking at Eqs.~\eqref{eq:VsqPPW}
and~\eqref{eq:4loff},~\eqref{eq:8loff} it is easy recognized that the following identities hold,
\begin{equation}
\left(M_{aa}^{ij}\right)^2 = \frac{f^2}{16}{\cal V}_{ij}~,~~~~~
\left(M_{88}^{ij}\right)^2 = \frac{f^2}{12}{\cal V}_{ij}~,
\end{equation}
where ${\cal V}_{ij}$ is the coefficient that multiplies the
gradient term in the action of the Goldstone boson $U(1)_A$.
Therefore the positivity of the matrix ${\cal V}_{ij}$ reflects to
the positivity of the Meissner tensor. Since we have shown that both
in the case of the BCC and of the FCC crystals ${\cal V}_{ij}$ is
indeed defined positive, a real value for all of the Meissner masses
follows at the leading order in $\Delta/\delta\mu$.

We may ask how the higher order corrections can modify the result expressed in
Eqs.~\eqref{eq:4loff},~\eqref{eq:8loff}. To this end it is enough to investigate on the typical ratio
$\Delta/\delta\mu$ in a crystalline LOFF state. Unfortunately, as already stressed in the previous
sections, the free energy (and the gap parameters) in a generic crystalline structure can be computed
only in some approximation. In Ref.~\cite{Casalbuoni:2004wm} a smearing over the cell has been
introduced: in this scheme it was found that for the BCC the ratio $\Delta/\delta\mu$ lies in the
interval $0.35 - 0.40$; for the FCC the interval is $0.27 - 0.30$ (see Tables I and II
of~\cite{Casalbuoni:2004wm}). Since the corrections to Eqs~\eqref{eq:4loff},~\eqref{eq:8loff} are of
order $\Delta^4/\delta\mu^4$, we see that the next-to-leading order terms are suppressed as
$\Delta^2/\delta\mu^2$ when compared to the leading order results. At the worst, one has
$\Delta^2/\delta\mu^2\simeq0.2$ for the BCC and $\Delta^2/\delta\mu^2\simeq0.1$ for the FCC. Therefore
we argue that the next-to-leading order terms should not change in a dramatic way the main result of
Eqs.~\eqref{eq:4loff},~\eqref{eq:8loff}, namely positive squared Meissner masses for all of the gluons
in a crystalline LOFF phase. Of course this hint should be supported by the calculation of the
next-to-leading order correction: unfortunately the computation of such terms in BCC and/or FCC is much
more involved than in the simple plane wave case. Another strategy would be the calculation using the
smearing approach~\cite{Casalbuoni:2004wm} instead of the GL expansion of the Meissner tensor. We will
come back to this problem in the future.

\subsection{A brief comment about Higgs stability of the LOFF phases}
Before concluding this section we briefly discuss how the Higgs
instability could arise in the LOFF phases of more than a single
plane-wave, following the treatment discussed
in~\cite{Giannakis:2006gg}. The Higgs (or amplitude) instability is
related to inhomogeneous variation of the amplitude of gap parameter
$\Delta$. It has been studied in great detail for the homogeneous
g2SC phase~\cite{Giannakis:2006gg,Giannakis:2006vt,Iida:2006df},
where it has been related to the existence of the Sarma instability.

In the LOFF phase, one can introduce a variation of the gap
parameter in Eq.~\eqref{eq:GapLPFFcc} by means of the replacement
$\Delta \rightarrow \Delta + H({\bm r})$ with $H({\bm r})$ real,
with Fourier transform given by $H({\bm k})$:
\begin{equation}
\Delta({\bm r})_{PPW} \rightarrow \sum_{a=1}^P e^{i2{\bm
q}_a\cdot{\bm r}} \left[\Delta + \int\frac{d^3{\bm
k}}{(2\pi)^3}~e^{-i{\bm k}\cdot{\bm r}}H({\bm k}) \right]~.
\end{equation}

Keeping into account fluctuations in the magnitude of $\Delta$ and
neglecting the phase fluctuations (i.e. the Goldstones) one can
write the grand potential, in the Gaussian approximation, as
\begin{equation}
\Omega = \Omega_0 + \frac{1}{2}\int\frac{d^3{\bm
k}}{(2\pi)^3}~H^*({\bm k})\frac{\partial^2\Omega}{\partial H^*({\bm
k})\partial H({\bm k})}H({\bm k})~, \label{eq:der}
\end{equation}
where $\Omega_0$ is the grand potential evaluated at the minimum and
the derivative, evaluated at the minimum of the free energy, is
nothing but the self-energy of the fluctuations
(see~\cite{Giannakis:2006vt} for more details). The momentum
integral in Eq.~\eqref{eq:der} can be divided {\em grosso modo} into
an integral over small momenta (with respect to ${\bm q}$) and an
integral over larger momenta. Following [HUANG], in the low momentum
region one can replace, in the thermodynamic limit,
\begin{equation}
\frac{\partial^2\Omega}{\partial H^*({\bm k})\partial H({\bm k})}
\rightarrow \frac{\partial^2\Omega}{\partial\Delta^2}~;
\end{equation}
the derivative on the r.h.s. of the above equation is positive
because the LOFF state is in a minimum of the grand potential.
Therefore a LOFF phase can not suffer Higgs instability, at least
for small momenta.

\section{Conclusions}
In this paper we have computed the low energy properties of the
Goldstone mode $\phi$ both in the gapped and in the gapless 2SC
phase of QCD. We stress that such a Goldstone is related to the
axial U(1) current and it is not one of the would-be Goldstones to
be eaten up after gauging of color. In the calculation presented in
Sec.~\ref{sec:EA} we introduce the $\phi$ as an external field. The
integration over the fermion fields in the functional integral
allows for the computation of the squared decay constant and of the
squared velocity of $\phi$. In particular, we find a negative
squared velocity in the gapless regime $\delta\mu > \Delta$. The
simple proportionality relation between $v^2$ and the squared
Meissner mass of the $8^{th}$ gluon shows that the two
instabilities, in the Goldstone and in the gluon sectors, are
related and the removal of the second is equivalent to the removal
of the first one.

Because of the wrong sign of the squared Goldstone velocity, the
gradient of the Goldstone field may take on a non-vanishing
expectation value in the vacuum, with consequent breaking of the
symmetry under rotations. We call this phase cur-g2SC phase.  The
Lagrangian for the fluctuations can be constructed in the cur-g2SC
phase and does not suffer of the squared-velocity sign problem.

Since the thermodynamics of the cur-g2SC is the same as that of the
one-plane-wave LOFF state, its free energy is higher than the energy
of a multiple-plane-wave phase. Motivated by this observation we
have computed the low energy parameters of the effective lagrangian
for the $U(1)_A$ Goldstone in the multiple-plane-wave LOFF phase
(PPW): we find that that such crystalline phases do not present
instabilities towards formation of currents. Moreover we have
calculated the Meissner masses of the gluons in the PPW phases,
generalizing the results of~\cite{Giannakis:2005vw}: we find
chromo-magnetic stability.

The other instability which could appear in in these LOFF phases is
the Higgs instability, which has been studied for the homogeneous
g2SC and related to the Sarma instability. We have argued that the
Higgs instability should be absent in the PPW LOFF phase, at least
for small momenta. A complete study of the amplitude instability
requires in addition, in the Gaussian approximation, the calculation
of the self-energy of $H({\bm k})$ for each value of the 3-momentum.

A multiple-plane-wave LOFF seems in conclusion to be one of the best
candidates for the ground state of a two flavor color
superconductor, beside the gluonic phases considered
in~\cite{Gorbar:2005rx}. A comparison of the free energies of the
neutral LOFF phases in the multiple plane wave state and of the
phases with condensed gluons is therefore crucial for a deeper
understanding of the phase diagram of two flavor QCD (in the single
plane wave case this study has been performed
in~\cite{Kiriyama:2006ui}).

Beside the comparison of the PPW state with the gluonic phases, a
natural prosecution of this work should be the extension to the
three flavor superconductive phases of QCD. In this context a lot of
work has been done in order to remove the chromo-magnetic
instability of the gapless CFL phase, see~\cite{Huang:2005pv}.
Moreover, the three flavor superconductive crystalline phase does
not seem to have such an instability, at least when the Meissner
masses are computed in the Ginzburg-Landau
approximation~\cite{Ciminale:2006sm}. It would be interesting to
study the Goldstone properties both in the kaon condensed and in the
three flavor LOFF phases of QCD, in order to see whether the
instability found here is present also in these cases, and if it is
completely removed once the gluon sector is cured.

\acknowledgements We thank H.~Abuki, M.~Ciminale, M.~Huang,
T.~Schafer for enlightening discussions. Moreover we thank
G.~Nardulli for the careful reading of the manuscript and for his
comments.

\appendix
\section{An attempt to match low and high energy theories}\label{sec:screening}
In the previous sections we have computed the squared decay constant $f^2$ and the squared velocity
$v^2$ of the Goldstone boson $\phi$, both in the gapped and in the gapless 2SC phases of QCD. The
results are in Eq.~\eqref{eq:Fsq2} and Eq.~\eqref{eq:vSq1}. In particular, we find a decreasing decay
constant as the ratio $\delta\mu/\Delta$ is increased in the gapless phase. Moreover, when $\delta\mu >
\Delta$ we find a negative squared velocity of $\phi$.

For the gapped 2SC and CFL phases it is possible to compute the low
energy parameters directly from the screening masses of the gluons,
or more generally of spin 1 gauge fields, see
e.g.~\cite{Son:1999cm,Schafer:2001za,Beane:2000ms,Manuel:2000wm}. We
wish to perform such a program also in this case, trying to relate
the decay constant and the squared velocity of $\phi$ found in the
previous section to the Debye and Meissner masses of a fictitious
gauge boson related to the $U(1)_A$ symmetry. In this section we
compute these masses.

To begin with we promote the global $U(1)_A$ symmetry to a local
one: this is done in the usual textbook way by the introduction of a
fictitious gauge field $W_\mu$ whose coupling to the fermions is
described via the covariant derivative $D_\mu =
\partial_\mu + i Q_A W_\mu$. Here $Q_A$ is the charge of the
fermions under $U(1)_A$, $Q_A=\pm 1$ respectively for left and right
handed quarks. The calculation of the static screening masses is
similar to that presented in the main text for the low energy
parameters of the Goldstone mode; therefore here we simply quote the
main results. In this case one has to evaluate the polarization
tensor $\Pi_{\mu\nu}$ of the gauge boson $W_\mu$. According to
Ref~\cite{Nardulli:2002ma} we define
\begin{equation}
m_D^2 = \Pi_{00}(q_0=0,{\bm q}\rightarrow0)~,~~~ m_M^2\delta_{ij} =
-\Pi_{ij}(q_0=0,{\bm q}\rightarrow0)~.
\end{equation}

First we consider the Debye mass. After some computation we are left
with the expression
\begin{equation}
m_D^2 = \frac{4\mu^2}{\pi^2}\Delta^2\int_{0}^{+\infty}
\!\!\!\frac{d\ell_\parallel}{(\ell_\parallel^2+\Delta^2)^{3/2}} +
M_{blue}^2~,
\end{equation}
where the first addendum on the r.h.s. is the contribution of the
paired red and green quarks, while the second addendum  is the
contribution of the unpaired blue quarks given by the standard
many-body result
\begin{equation}
M_{blue}^2 = 2\times\frac{\mu^2}{\pi^2} ~.
\end{equation}
In the gapless phase $\delta\mu
> \Delta$ it is interesting to distinguish, in the loop integral involving paired quarks,
the contribution of the quarks whose loop momentum $\ell_\parallel$
lies in the blocking region~\eqref{eq:BRdef} from the contribution
of the quarks living in the pairing region (defined as the
complementary of the blocking region):
\begin{eqnarray}
\int_{0}^{+\infty}
\!\!\!\frac{d\ell_\parallel}{(\ell_\parallel^2+\Delta^2)^{3/2}} &=&
\int_{0}^{\sqrt{\delta\mu^2-\Delta^2}}
\!\!\!\frac{d\ell_\parallel}{(\ell_\parallel^2+\Delta^2)^{3/2}} +
\int_{\sqrt{\delta\mu^2-\Delta^2}}^{+\infty}
\!\!\!\frac{d\ell_\parallel}{(\ell_\parallel^2+\Delta^2)^{3/2}}
\nonumber \\
&=&\frac{\sqrt{\delta\mu^2-\Delta^2}}{\Delta^2\delta\mu} +
\left(\frac{1}{\Delta^2}-\frac{\sqrt{\delta\mu^2-\Delta^2}}{\Delta^2\delta\mu}\right)
\nonumber \\
&\equiv& M^2_{BR} +  M^2_{PR}~,\label{eq:Div}
\end{eqnarray}
where $BR$ and $PR$ stem respectively for Blocking Region and
Pairing Region. With these definitions at hand one can write
\begin{equation}
m_D^2 = M^2_{BR} +  M^2_{PR} + M^2_{blue} =
6\times\frac{\mu^2}{\pi^2}~. \label{eq:DebyeMass}
\end{equation}
The Debye mass is independent of the value of the ratio $\delta\mu/\Delta$ and is equal to the result
of the normal phase. This can be understood by noticing that both the condensate and the fermion fields
have $U(1)_A$ charge, so one expects screening both in the gapped and in the gapless phase.

Next we turn to the Meissner mass. We get
\begin{equation}
m_M^2 =\frac{4\mu^2}{3\pi^2}\left(1-
\theta(\delta\mu-\Delta)\frac{\delta\mu}{\sqrt{\delta\mu^2-\Delta^2}}\right)~.\label{eq:MeissnerMass}
\end{equation}
In the gapped 2SC phase one has $\delta\mu<\Delta$ and the squared
Meissner mass is positive and constant. On the other hand in the
gapless phase $\delta\mu>\Delta$ the squared Meissner mass is
negative and divergent ad the transition point. The divergence is
related to the divergence of the density of gapless quasi-particle
states.

The calculations of the Debye and Meissner masses which lead to the results in
Eqs.~\eqref{eq:DebyeMass} and~\eqref{eq:MeissnerMass} are done using the so called high energy
theory~\cite{Beane:2000ms}, that is using a lagrangian written in terms of the quarks degrees of
freedom only. It would be interesting to make the same calculation in the so called low energy theory,
that is by using a lagrangian written in terms of the light degrees of freedom, the Goldstone mode and
the gapless quarks. The matching of the results obtained by the two procedures should allow to compute
$f^2$ and $v^2$.

In the gapped CFL phase of QCD the matching procedure is easy to
perform since the light degrees of freedom are the Goldstone bosons
(an octet plus a singlet), while the high energy theory is defined
in terms of the nine gapped quarks. Also in the 2SC phase the paired
quarks are gapped, so the low energy degrees of freedom are the
singlet $\phi$ and the unpaired blue quarks. In the gapless 2SC
phase, on the other hand, one has also gapless paired fermions in
the low energy spectrum, so one should consider them in the matching
procedure.

Let us consider first the case of the gapped 2SC phase. On the basis
of invariance under $U(1)_A$ it is easy to recognize that the
effective lagrangian for $U = \exp(i \phi/f)$ can be cast in the
form
\begin{equation}
{\cal L} = \frac{f^2}{2}\left(\partial_0 U \partial_0 U^* - v^2
\bm{\nabla} U \bm{\nabla} U^*\right)\label{eq:LagrEta}
\end{equation}
(with this definition the kinetic term of the field $\phi$ gets the canonical normalization). The
gauging of the $U(1)_A$ is made by the replacement of the usual derivative with the covariant one
$D_\mu =
\partial_\mu + i W_\mu$. After the gauging, a mass term for the
fictitious gauge field $W_\mu$ is obtained, namely
\begin{equation}
\delta{\cal L} = \frac{f^2}{2}\left(W_0^2 - v^2 {\bm W}\cdot {\bm
W}\right) \equiv \frac{M_D^2}{2} W_0^2 - \frac{M_M^2}{2}{\bm W}\cdot
{\bm W}~,\label{eq:massesLET}
\end{equation}
where we denote by the capital letter $M_{D,M}$ the contribution of
the Goldstone to the Debye and the Meissner mass of $W_\mu$. However
one has to add the contribution of the blue unpaired quarks to
Eq.~\eqref{eq:LagrEta} in order to properly describe the low energy
theory. The blue quarks contribute only to the Debye mass, as they
are not superconductive and do not screen static ``magnetic'' fields
${\bm W}$. As a consequence, the Debye and Meissner masses in the
low energy theory are
\begin{equation}
m_D^2 =  M_D^2 + M^2_{blue}~,~~~m_M^2 =  M_M^2~.\label{eq:com1}
\end{equation}
On the other hand one can read $m_D^2$ and $m_M^2$ from the high
energy calculations, see Eqs.~\eqref{eq:DebyeMass} and
~\eqref{eq:MeissnerMass}. Comparison allows to compute $f^2$ and
$v^2$,
\begin{equation}
f^2_{2SC} = \frac{4\mu^2}{\pi^2}~,~~~v^2_{2SC} = \frac{1}{3}~.
\end{equation}

Now we turn to the gapless 2SC phase and consider the Debye mass. In
this case the effective lagrangian for $\phi$ has the same form of
the gapped phase, Eq.~\eqref{eq:LagrEta}, since it is related only
to the symmetries of the ground state  and not to the
absence/presence of gapless excitations in the spectrum. Therefore
the contribution of $\phi$ to the screening masses is described by
Eq.~\eqref{eq:massesLET} and $m_{D}^2$ is given in the low energy
theory by
\begin{equation}
m_D^2 = M_D^2 + M^2_{blue} + M^2_{gapless}~,
\end{equation}
where $M^2_{gapless}$ is the (till unknown) contribution of the
gapless fermions. A direct comparison with Eq.~\eqref{eq:DebyeMass}
allows the identification
\begin{equation}
f^2  = \frac{4\mu^2}{\pi^2} - M^2_{gapless}~. \label{eq:Match1}
\end{equation}

At this point, if we knew $M^2_{gapless}$ we should be able to
determine $f^2$ in the low energy theory. However this contribution
to the Debye mass is not well defined, as to calculate it one should
choose an appropriate infrared cutoff and integrate out all quarks
whose momenta are above such a cutoff. In this way, the theory would
contain only the light degrees of freedom and the computation of
$M^2_{gapless}$ would be straightforward.

However, in the gapless regime we can use a different strategy: we
already have computed $f^2$ using a completely different approach in
the previous section, see Eq.~\eqref{eq:Fsq2}. Using this result,
the matching condition~\eqref{eq:Match1} can be used {\em not} to
compute $f^2$ {\em but} to evaluate $M^2_{gapless}$, namely
\begin{equation}
M^2_{gapless} = \frac{4\mu^2}{\pi^2}\theta(\delta\mu-\Delta)
\frac{\sqrt{\delta\mu^2-\Delta^2}}{\delta\mu}~.\label{eq:Match2}
\end{equation}
This result is equal to $M_{BR}^2$ in  Eq.~\eqref{eq:Div}, which is
obtained by integrating in the loop over the quark momenta in the
blocking region. This coincidence seems to show that the g2SC phase
can be described like a two-component Fermi liquid, with a normal
part given by the blocking region and relevant for the low energy
dynamics, and a superfluid part living in the pairing region.

We consider now the Meissner mass: this case is instructive since it
shows that the aforementioned interpretation of the g2SC as a
two-component superfluid leads to meaningful results once one tries
the matching. Following our interpretation we argue that the quarks
in the blocking region, as they behave as a normal liquid, do not
contribute to the Meissner mass. The same is true for the unpaired
blue quarks. Finally, the quarks in the pairing regions are
superfluid and do not contribute to the low energy dynamics. As a
consequence, the only contribution to the Meissner mass in the low
energy theory is given by the $\phi$ as in the 2SC phase, see
Eqs.~\eqref{eq:massesLET} and~\eqref{eq:com1}. Matching with the
high energy result in Eq.~\eqref{eq:MeissnerMass} one gets
\begin{equation}
v^2 = \frac{m_M^2}{f^2} = \frac{m_M^2}{m_D^2-
M^2_{gapless}-M^2_{blue}}~,
\end{equation}
which leads to the result quoted in Eq.~\eqref{eq:vSq1} once
Eqs.~\eqref{eq:DebyeMass},~\eqref{eq:MeissnerMass}
and~\eqref{eq:Match2} are used.

\end{document}